\documentclass[aps,prb,floatfix,twocolumn,showpacs]{revtex4}
\usepackage{epsfig,amsmath,amssymb,rotating,multirow}

\begin{document}

\hsize\textwidth\columnwidth\hsize\csname@twocolumnfalse\endcsname

\title{Spin-orbit coupling and anisotropic exchange in two-electron
double quantum dots  }

\author{Fabio Baruffa$^1$, Peter Stano$^{2,3}$ and Jaroslav Fabian$^1$}
\affiliation{$^1$Institute for Theoretical
Physics, University of Regensburg, 93040 Regensburg, Germany\\
$^2$ Institute of Physics, Slovak Academy of Sciences, Bratislava 845 11, Slovakia\\
$^3$Physics Department, University of Arizona, 1118 East 4th Street, Tucson,
Arizona 85721, USA}

\vskip1.5truecm
\begin{abstract}
The influence of the spin-orbit interactions on the energy spectrum of
two-electron laterally coupled quantum dots is investigated. The effective
Hamiltonian for a spin qubit pair proposed in F.Baruffa et al., Phys. Rev. Lett. 104, 126401 (2010) is
confronted with exact numerical results in single and double quantum dots in
zero and finite magnetic field. The anisotropic exchange Hamiltonian is found
quantitatively reliable in double dots in general. There are two
findings of particular practical importance: i) The model stays valid even
for maximal possible interdot coupling (a single dot), due to the absence of
a coupling to the nearest excited level, a fact following from the dot
symmetry. ii) In a weak coupling regime, the Heitler-London approximation
gives quantitatively correct anisotropic exchange parameters even in a finite
magnetic field, although this method is known to fail for the isotropic
exchange. The small discrepancy between the analytical model (which employs
the linear Dresselhaus and Bychkov-Rashba spin-orbit terms) and the
numerical data for GaAs quantum dots is found to be mostly due to the cubic Dresselhaus term.
\end{abstract}
\maketitle

\section{Introduction}

The lowest singlet and triplet states of a two electron system are split by the exchange energy. This is a direct consequence of the Pauli exclusion principle and the Coulomb interaction. As a result, a spin structure may appear even without explicit spin dependent 
interactions.\cite{Lieb:PR1962} 

In quantum dot spin qubits\cite{Hanson:RMP2007} the exchange interaction implements a fundamental two qubit gate.\cite{Loss:PRA1998,Hu:PRA2000} Compared to single qubit gates,\cite{Nowack:Science2007,Koppens:Nature2006} the exchange-based gates are much faster\cite{Petta:Science2005} and easier to control locally, motivating the solely exchange-based quantum computation.\cite{Coish:PRB2007} The control is based on the exponential sensitivity of the exchange energy on the inter-particle distance. Manipulation then can proceed, for example, by shifting the single particle states electrically
\cite{Popsueva:PRB2007,Petta:Science2005,Laird:PRL2006} or compressing them magnetically.
\cite{Burkard:PRB2000} 

The practical manipulation schemes require quantitative knowledge of the exchange energy. The configuration interaction,\cite{Merkt:PRB1991,DeSousa:PRA2001,Dybalski:PRB2005,Pedersen:PRB2007,
Climente:PRB2007} a numerically exact treatment, serves as the benchmark for usually adopted approximations. The simplest one is the Heitler-London ansatz, in which one particle in the orbital ground state per dot is considered. The exchange asymptotic in this model differs from the exact\cite{Gorkov2:PRB2003,Melnikov:PRB2006} and the method fails completely in finite magnetic fields. Extensions of the single particle basis include the Hund-Mullikan,
\cite{Burkard:PRB2000} Molecular Orbital,\cite{DeSousa:PRA2001,Nguyen:PRB2008} or Variational method.\cite{Kandemir:PRB2005,Dybalski:PRB2005} Other approaches, such as the Hartree-Fock,\cite{Pfannkuche:PRB1993,Yannouleas:PRB2003,Yannouleas:IJQC2002} random phase approximation\cite{Serra:PRB2003} and (spin-)density functional theory\cite{Saarikoski:EPJB2001} were also examined. None of them, however, is reliable in all important regimes,\cite{He:PRB2005,Pedersen:PRB2007,Hu:PRA2000} which include weak/strong interdot couplings, zero/finite magnetic field and symmetric/biased dot. 

The spin-orbit interaction, a non-magnetic spintronics workhorse,\cite{Zutic:RMP2004} is a generic feature in semiconductor quantum dots.\cite{Fabian:APS2007}
Although it is usually weak, it may turn out of major importance as, 
for example, for the spin relaxation,
\cite{Stano:PRL2006,Semenov:PRB2007,Meunier:PRL2007,Sasaki:PRL2005,Jiang:PRB2008,Badescu:PRB2005,Cheng:PRB2004,Shen:PRB2007,Climente:PhE2008,Climente:PRB2007,MezaMontes:PRB2008,
Pfund:PRB2009}, or, more positively, a 
handle for the electrical spin manipulation.\cite{Khomitsky:PRB2009,Stano:PRB-77-045310} It is natural to expect that the presence of the spin-orbit interaction will influence the exchange Hamiltonian.\cite{SanJose:PRB2008} The resulting corrections to the rotationally symmetric exchange Hamiltonian are  referred to as the anisotropic exchange (we do not consider other sources than the spin-orbit interaction\cite{Badescu:PRB2005,Imamura:PRB2004,Yang:PRB2006,Glazov:PRB2009}). Stringent requirements of the quantum computation algorithms motivate studies of the consequences of the anisotropic exchange of a general form on quantum gates.
\cite{Devitt:PRA2006,Chutia:PRB2006,Zhao:PRB2006} Usually, the anisotropic exchange is viewed as a nuisance to be minimized.\cite{Stepanenko:PRL2004,Bonesteel:PRL2001,Burkard:PRL2002} On the other hand, it was considered as a possible way of implementing the quantum gates.
\cite{Stepanenko:PRL2004,Wu:PRA2002} In both views, it is of utter importance to know the strength and the form of the anisotropic exchange. Since the spin-orbit interaction is weak, it is enough to answer the following question: What is the anisotropic exchange in the leading order?

Surprisingly, arriving at the answer was not straightforward at all. The Dzyaloshinskii-Moriya\cite{Moriya:PR1960,Dzyaloshinsky:JPCS1958} interaction is of the first order in spin-orbit coupling. However, since it couples only states split by the isotropic exchange, it is necessary to consider also the second order anisotropic exchange terms to arrive at correct energies.
\cite{Gangadharaiah:PRL2008,Shekhtman:PRL1992,Zheludev:PRB1999} Ref.\onlinecite{Kavokin:PRB2001} suggested such a Hamiltonian, which was unitarily equivalent to the isotropic exchange Hamiltonian, with the exchange energy renormalized in the second order. This was later revisited,\cite{Kavokin:PRB2004,Gorkov:PRB2003} with the following conclusion: In zero magnetic field, the two qubit Hamiltonian is, up to the second order in the linear-in-momenta spin-orbit interaction, unitarily equivalent to the isotropic exchange Hamiltonian
in the weak coupling limit, with the unchanged exchange energy. Further corrections appear in the third order. In the unitary operator providing the change of the basis, the spin-orbit interaction appears in the linear order. These results are a consequence of the special form of the spin-orbit interaction, which in the leading order leads to a spatially dependent spin rotation.\cite{Levitov:PRB2003}

In the short version of this article,\cite{Baruffa:PRL2010} we derived the leading order anisotropic exchange terms which appear in a finite magnetic field. We derived 
all anisotropic exchange parameters in a form 
valid for arbitrary interdot coupling. We also compared the results obtained using the first order versus the second order treatment of the spin-orbit interactions. The main goal of the present work is a detailed assessment of the quantitative reliability of the presented anisotropic exchange model comparing with exact numerical results. Specifically, we examine the model in the strong and weak coupling regimes [corresponding to single (Sec.~\ref{singledot}) 
and double (Sec.~\ref{doubledot}) dots, respectively] and in zero and finite perpendicular magnetic field. We also study the role of the cubic Dresselhaus term 
(Sec.~\ref{cubicdresselhaus}), whose action does not correspond to a spatial texture (in the leading order) and could potentially become dominant over the linear terms, changing the picture considerably. In addition to that, we supply the derivations, not presented in the short version 
(Sec.~\ref{derivation}) and a detailed account of our numerical method (App.~\ref{appendix}).

The analytical pitfalls in evaluating the isotropic exchange are well known.
\cite{Gorkov2:PRB2003,Herring:RVM1962} On  top of that, the anisotropic exchange is a (very) small correction to the exponentially sensitive isotropic exchange, and therefore it is involved to extract even numerically. Our main conclusion here is that the presented analytical model is valid in all studied regimes. Quantitatively, the effective parameters are usually within a factor of 2 from their counterparts derived from the numerically exact spectra. The main source of the discrepancy is the cubic Dresselhaus term. Surprisingly, in the most important regime for quantum dot spin qubits, namely the weak coupling, the Heitler-London approximation works great for the anisotropic exchange, even though it fails badly for the isotropic one. This finding justifies  using simple analytical formulas for the anisotropic exchange parameters.

\section{Model}
Our system is a two-dimensional electron gas confined in a [001] plane of
a zinc-blende semiconductor heterostructure. An additional lateral potential
with parabolic shape defines the double quantum dot.
We work in the single band effective mass  approximation.
The two-electron Hamiltonian is a sum of the orbital part and the spin dependent part,
\begin{equation}\label{ham_tot}
H_{tot} = H_{orb} + \sum_{i=1,2} H_{so,i} + H_{Z,i} = H_{orb}+H_{so}+H_Z,
\end{equation}
where the subscript $i$ labels the two electrons. The orbital Hamiltonian is
\begin{equation}\label{ham_orb}
H_{orb}=\sum_{i=1,2} \left(T_{i}+V_{i}\right) + H_C.
\end{equation}
Here, $T_i=\hbar^2{\bf K}_i^{2}/2m$ is 
the kinetic energy with the effective mass $m$
and the kinetic momentum 
$\hbar {\bf K}_i = \hbar {\bf k}_i + e {\bf A}_i =
-i\hbar{\boldsymbol \nabla}_i+e {\bf A}_i$;
$e$ is the proton charge and ${\bf A}_i = B_z/2(-y_i,x_i)$ is the vector potential of the magnetic field ${\bf B} = (B_x,B_y,B_z)$.
The potential $V$ describes the quantum dot geometry
\begin{equation}\label{confinement}
V_i=\frac{1}{2}m\omega_0^2 \mbox{min} \{({\bf r}_i-{\bf d})^2,
({\bf r}_i+{\bf d})^2\}.
\end{equation}
Here $l_0=(\hbar/m\omega_0)^{1/2}$
is the confinement length, $2d$ measures the distance between the two
potential minima, the vector ${\bf d}$ defines the main dot axis 
with respect to the crystallographic axes and
$E_0=\hbar\omega_0$ is the confinement energy.
The Coulomb interaction between the two electrons is
\begin{equation}\label{coulomb}
H_C = \frac{e^2}{4\pi\epsilon_0\epsilon_r}\frac{1}{|{\bf r}_{1}-{\bf r}_{2}|},
\end{equation}
where $\epsilon_0$ is the vacuum dielectric constant and $\epsilon_r$ is the
dielectric constant of the material.

The lack of the spatial inversion symmetry 
is accompanied by the spin-orbit  interaction of a general form
\begin{equation}\label{spinorbit}
H_{so,i} = {\bf w}_i \cdot \boldsymbol{ \sigma }_i,
\end{equation}
where the vector ${\bf w}$ is kinetic momentum dependent. In the
semiconductor heterostructure, there are two types
of spin-orbit interactions.
The Dresselhaus spin-orbit
interaction, due to the bulk inversion asymmetry of the zinc-blende structure, 
consists of two terms, one linear and one
cubic in momentum\cite{Fabian:APS2007}
\begin{eqnarray}\label{dresselhaus}
{\bf w}_{D,i} &=& \gamma_c\langle K_{z,i}^2\rangle\left(-K_{x,i},K_{y,i},0\right),\\
{\bf w}_{D3,i} &=& \gamma_c/2\left(K_{x,i}K^{2}_{y,i}, -K_{y,i}K^{2}_{x,i},0\right) + \mbox{H.c.},
\end{eqnarray}
here H.c. denotes the Hermitian conjugate.
The interaction strength $\gamma_c$ is a material parameter, the angular brackets in ${\bf w}_{D}$
denote the quantum averaging in the ${\bf z}$ direction. Since both electrons are in the ground state of the perpendicular confinement, we have $ \langle K_{z,1}^2\rangle = \langle K_{z,2}^2\rangle = \langle K_z^2\rangle$, the value depending on the confinement details.
A confinement asymmetry along the
growth direction (here ${\bf z}$) gives rise to the
Bychkov-Rashba term\cite{Fabian:APS2007}
\begin{equation}\label{rashba}
{\bf w}_{BR,i} = \alpha_{BR}\left(K_{y,i},-K_{x,i},0\right).
\end{equation}
The coupling $\alpha_{BR}$ of the interaction is structure dependent
and can be, to some extent, experimentally
modulated by the top gates potential.
Equations~(\ref{dresselhaus}-\ref{rashba}) are valid for a coordinate system where the x and y axes are chosen along 
[100] and [010] directions, respectively.
Below we use the effective
spin-orbit lengths defined as $l_{br}=\hbar^2/2m\alpha_{BR}$
and $l_{d}=\hbar^2/2m\gamma_c\langle K_z^2\rangle$.

The spin is coupled to the magnetic field through the Zeeman interaction
\begin{equation}\label{zeeman}
H_{Z,i} = \frac{g}{2}\mu_B {\bf B}\cdot 
\boldsymbol{ \sigma }_i = \mu {\bf B}\cdot \boldsymbol{ \sigma }_i,
\end{equation}
where $g$ is the effective gyromagnetic factor, $\mu_B=e\hbar/2m_e$ 
is the Bohr magneton (alternatively, we use a renormalized 
magnetic moment $\mu$) and
$\boldsymbol{ \sigma } $ is the vector of the Pauli matrices.

In lateral quantum dots the Coulomb energy $E_C$ is comparable
to the confinement energy and the correlation between the electrons strongly influence
the states.\cite{Lucignano:PRB2004,Rontani:SSC2001} One can compare the energies considering
\begin{equation}
\frac{E_C}{E_0}=\frac{e^2}{4\pi\epsilon_0\epsilon_r}\langle r^{-1} \rangle\frac{m l_0^2}{\hbar^2}\sim\frac{l_0^2}{l_C \langle r \rangle},
\end{equation}
where the Coulomb length 
$l_C=e^2 m/4\pi \epsilon_0 \epsilon_r \hbar^2$ is a material parameter
and $\langle r \rangle$ is the mean distance between the electrons.
In GaAs $l_C\approx 10$ nm, while a typical lateral dot has $l_0\approx 30$ nm,
corresponding to $E_0\approx1$ meV.
The mean length $\langle r \rangle$ is of the order of the confinement length, 
if the two electrons are on the same dot, and of the 
interdot distance, if the electrons are on different dots.
In the first case, the Coulomb energy is typically $3$ meV. 
In the second case (one electron per dot) the Coulomb interaction is 
typically at least $1$ meV.

The strength of the Coulomb interaction precludes 
the use of perturbative methods. Therefore, to
diagonalize the two electron Hamiltonian
Eq.~(\ref{ham_tot}), we use the exact numerical treatment,
the Configuration interaction method. Details are given
in App.~\ref{appendix}. Below we consistently use the notation of $\Phi$
for spinor and $\Psi$ for orbital wavefunctions. They fulfill
the equations  $H_{tot}\Phi=E\Phi$ and $H_{orb}\Psi=E\Psi$, respectively.

We use the GaAs realistic parameters: $m=0.067m_e$
($m_e$ is the free electron mass), $g=-0.44$, $\epsilon_r=12.9$ and 
$\gamma_c = 27.5\mbox{ eV\AA{}}^3$. The coupling of the linear Dresselhaus term
is $\gamma_c\langle K_z^2 \rangle = 4.5\mbox{ meV\AA{}}$
and of the Bychkov-Rashba term is $\alpha_{BR} = 3.3 \mbox{ meV\AA{}}$,
corresponding to the effective spin-orbit lengths $l_d=1.26\mu$m
and $l_{br}=1.72\mu$m, 
according to the recent experiments.\cite{Stano:PRB2005,Stano:PRL2006} We use the confinement
energy $\hbar\omega_0=1.1\mbox{ meV}$, which corresponds to the confining length
$l_0 = 32 \mbox{ nm}$, in line with an experiment.\cite{Elzerman:Nature2004}

\subsection{Unitarily transformed Hamiltonian}

Analytically, we will analyze
the role of the spin-orbit interactions in the two-electron spectrum
using the perturbation theory. This approach is appropriate since
the spin-orbit energy corrections are small compared to the typical confinement energy. 
For a GaAs quantum dot the ratio between the confinement length and the  
spin-orbit length $l_0/l_{so} \sim 10^{-2} \div  10^{-3}$. Furthermore,
for a magnetic field of $1$ Tesla, the ratio between the Zeeman energy 
and the confinement energy is $\mu B/E_0 \sim 10^{-2}$.
Therefore the spin-orbit interactions are small perturbations, comparable in strength to the Zeeman term at $B=1$ Tesla.

We consider the perturbative solution of the Hamiltonian Eq.~(\ref{ham_tot}). 
We transform the Hamiltonian to gauge out the linear spin-orbit terms,
\cite{PRL-87-256801, Levitov:PRB2003}
(we neglect the cubic Dresselhaus term in the analytical models) 
\begin{equation}\label{eq:unit_transf}
 H_{tot} \rightarrow U H_{tot} U^\dag = H_{orb} + H_{Z} + 
\overline{H}_{so},
\end{equation}
using the operator
\begin{equation}\label{unitary_operator}
 U = \exp\left(-\frac{\rm i}{2} {\bf n}_{1}\cdot{\boldsymbol \sigma}_{1}
-\frac{\rm i}{2} {\bf n}_{2}\cdot{\boldsymbol \sigma}_{2}\right),
\end{equation}
where
\begin{equation}\label{direction}
{\bf n}_i = \left(\frac{x_i}{l_d}-\frac{y_i}{l_{br}},
\frac{x_i}{l_{br}}-\frac{y_i}{l_{d}},0 \right).
\end{equation}
Keeping only terms up to the second order in the spin-orbit  and  Zeeman couplings, we get the
following effective spin-orbit interactions 
$\overline{H}_{so} = H_{so}^{(2)}+H_{Z}^{(2)}$, where
\begin{eqnarray}
H_{so}^{(2)} = \sum_{i=1,2}\left( -K_+ + K_- L_{z,i}\sigma_{z,i}/\hbar \right),
\label{eq:hsoprime}\\
H_{Z}^{(2)} = \sum_{i=1,2} -(\mu{\bf B}\times {\bf n}_{i})\cdot 
 \boldsymbol{ \sigma }_{i}.\label{eq:hzprime}
\end{eqnarray}
Here, $L_{z,i}/\hbar = x_i K_{y,i}-y_i K_{x,i}$, and
\begin{eqnarray}\label{eq:Kpm}
K_{\pm} = \left(\frac{\hbar^2}{4ml_{d}^2} \pm \frac{\hbar^2}{4ml_{br}^2}\right).
\end{eqnarray}
Equation~(\ref{eq:hzprime}) describes the mixing between the Zeeman
and spin-orbit interactions, which is linear in the spin-orbit couplings.
It disappears in zero magnetic field, where only the terms in 
Eq.~(\ref{eq:hsoprime}) survive -- 
a sum of an overall constant shift of $2K_+$
and the spin-angular momentum operators. Both of these are
quadratic in the spin-orbit couplings. 

The point of the transformation, which changes the form of the spin-orbit interactions, is that the transformed interactions are much weaker (being the second, instead of the first order in the spin-orbit/Zeeman couplings). Of course, both Hamiltonians are equivalent, giving the same exact energies. However, a perturbative expansion of the transformed Hamiltonian converges much faster.

\subsection{Orbital functions symmetry}

The symmetry of the two-electron wavefunctions $\Psi$ 
has important consequences, for example, in the form of
selection rules for the couplings 
between the states due to the spin-orbit interactions.
The choice of the potential in Eq.~(\ref{confinement}) is motivated
by the fact that
for small ($d\rightarrow0$) and large ($d\rightarrow\infty$)
interdot distance the eigenstates of the single particle Hamiltonian
converge to the single dot solutions centered
at $d=0$ and $x=\pm2d$, respectively. 
For zero magnetic field,
since the double dot potential does not have the rotational symmetry around
the $z$ axis, the inversions of the coordinate
along axes of the confinement potential ($x$ and $y$) are the
symmetries involved. Indeed, the orbital Hamiltonian Eq.~(\ref{ham_orb})
commutes with the inversion operator $I_x$ 
and $I_y$, $[H_{orb},I_{x,y}]=0$. Furthermore $[H_{orb},I]=0$, where 
$I=I_x I_y$ is the inversion of both
axes simultaneously.
All these operations belong to the $C_{2v}$ group.
Accordingly, the wavefunctions transform as the functions 1, x, xy, and y, which
represent this group.
If a perpendicular magnetic field is applied, 
only the total inversion operation, $I=I_x I_y$, commutes with
the Hamiltonian  and the wavefunction is symmetric or
antisymmetric with respect to the total inversion -- this is due to the lack 
of $I_x$ and $I_y$ symmetry of 
the kinetic energy operator.  
The Slater determinants (the two-electron basis that we use in the
diagonalization procedure -- see App.~\ref{appendix}) have also definite symmetries,
if they are built from single particle states of definite symmetry 
(see App.~\ref{SDsymmetries}). 

We define the functions $\Psi_{\pm}$ to be the lowest eigenstates of 
the orbital part of the Hamiltonian, $H_{orb}\Psi_{\pm} = E_\pm \Psi_{\pm}$ with
the following symmetry,
\begin{equation}\label{eq:part_ex}
P \Psi_{\pm} = \pm \Psi_{\pm},
\end{equation}
where $Pf_{1}g_{2} = f_{2}g_{1}$ is the particle exchange
operator.
We observe that $\Psi_{\pm}$ have, in addition to the particle exchange symmetry,
also a definite spatial symmetry. In further
we assume they fulfill  
\begin{equation}\label{eq:symmetry1}
I_{1} I_{2} \Psi_\pm = \pm \Psi_\pm.
\end{equation}
We point out that while Eq.~(\ref{eq:part_ex}) is a definition, 
Eq.~(\ref{eq:symmetry1}) is an assumption based on an observation. In zero
magnetic field $I_{1} I_{2} \Psi_+ = + \Psi_+,$ follows from the
Mattis-Lieb theorem.\cite{Lieb:PR1962} 
For the validity of Eq.~(\ref{eq:symmetry1}) we resort to numerics---we saw it to hold in all cases we studied.

\begin{figure}
\centerline{\psfig{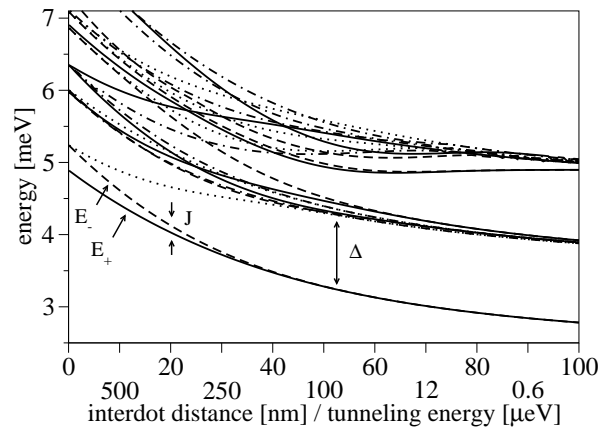}}
\caption{Two-electron energy spectrum of a double dot at zero magnetic field
as a function of the interdot distance and the tunneling energy. The spatial
symmetries of wavefunctions, 1, x, xy, and y are denoted as solid,
dashed, dotted-dashed and dotted line, respectively. The two lowest energies are labeled;
they are split by the isotropic
exchange energy $J$. The energy separation between the lowest states
and the higher exited states is denoted by $\Delta$.}
\label{fig:2dot_spectrum}
\end{figure}

Figure~\ref{fig:2dot_spectrum}
shows the calculated double dot spectrum at zero magnetic field
without the spin-orbit interactions. 
The two lowest states $\Psi_{\pm}$ 
are split by the exchange energy $J$.
In the single dot case ($d=0$), the ground state is non degenerate, while
the first exited state is doubly degenerate. Increasing the interdot distance,
this degeneracy is removed, as the two states
have different spatial symmetry (x and y).  
The energy of the states $\Psi_{\pm}$ is separated from the higher states 
by an  energy gap $\Delta$. This gap allows us to consider only the two lowest
orbital states when studying the spin-orbit influence on the lowest
part of the two-electron spectrum. 
Indeed, in the double dot $\Delta$ is of the order of $1$ meV, while
the spin-orbit interactions are two orders of magnitude smaller.
In the case of $\Delta=0$, the two orbital states approximation
can be improved including more states (although we show below this is not in fact necessary for a qubit pair in a circular dot).

Without the spin-orbit interactions,
the eigenstates of the  Hamiltonian Eq.~(\ref{ham_tot})
are separable in  the spin and orbital degrees of freedom. We get the four lowest 
states
 by supplementing $\Psi_{\pm}$ with spinors,
forming the singlet and triplets:
\begin{equation}\label{eq:basis4}
\{ \Phi_i \}_{i=1,\ldots,4}=\{ \Psi_+ S, \Psi_- T_0, \Psi_- T_+, \Psi_- T_-  \}.
\end{equation}
Here $S=1/\sqrt{2}(\left|\uparrow\downarrow\right>-\left|\downarrow\uparrow\rangle\right>)$ 
is a singlet spinor built out of two spin-1/2 spinors, 
$T_0=1/\sqrt{2}(\left|\uparrow\downarrow\right>+\left|\downarrow\uparrow\right>)$,
$T_+=\left|\uparrow\uparrow\right>$, $T_-=\left|\downarrow\downarrow\right>$
are the three possible triplets; the quantization axis
is chosen along the magnetic field. 

The symmetry  leads to  selection rules for the matrix elements
between  two electron states. In zero perpendicular magnetic field,
because the $L_z$ operator transforms as xy, 
the singlet
and triplets are not coupled, up to the second order
in the spin-orbit interactions,  
$\langle\Phi_1|\overline{H}_{so}|\Phi_{2,3,4}\rangle=0$. 
The only contribution is due to the constant $K_{+}$.
For non zero perpendicular
magnetic field, the singlet and a triplet are coupled only
if their orbital parts have the opposite spatial symmetry, due to the
term in Eq.~(\ref{eq:hzprime}). 
The non-vanishing matrix elements are listed in Table~\ref{table 1}.
\begin{table}[htbp]
\begin{center}
\begin{tabular}{c|c|c}
\hline
\hline
$\hat{O}$ & zero perpendicular field & finite perpendicular field \\
\hline
$L_{z,1}$ & never & $j_1 = j_2$ \\
\hline
$\bf{n_1}$ & $j_1 \neq j_2$ & $j_1 \neq j_2$ \\
\hline
\hline
\end{tabular}
\end{center}
\caption{
Conditions on the orbital symmetries for the matrix elements $\langle\Psi_{1}|\hat{O}|\Psi_2\rangle$ to be non-zero.
 The orbital symmetries are defined by 
$I\Psi_{1,2} = j_{1,2}\Psi_{1,2}$. }
\label{table 1}
\end{table}

\subsection{Effective Hamiltonians}\label{derivation}

Here we derive effective four level Hamiltonians, which provide understanding for
the numerical results. We follow two different approaches:
(i) restriction of the total Hamiltonian, Eq.~\eqref{ham_tot}, to the
 basis in Eq.~(\ref{eq:basis4}); (ii) including higher excited states through
a sum rule using the Schrieffer-Wolff transformation with the unitary
operator, Eq.~\eqref{unitary_operator}. Then we compare
the two models, including their simplifications using 
the Heitler-London approximation, 
to demonstrate the quality of their description of the two-qubit subspace.

We restrict the Hilbert space of the double dot to the four lowest
functions Eq.~(\ref{eq:basis4}) to describe the qubit pair.
We start with the case of zero spin-orbit interactions. 
In the external magnetic field, the two triplets $T_+$
and $T_-$ are split by twice the Zeeman energy $E_Z=2\mu B_z$. 
The restriction of the Hamiltonian Eq.~(\ref{ham_tot}) to the basis 
Eq.~(\ref{eq:basis4}) produces a diagonal matrix 
\begin{equation}\label{eq:4x4}
H_{iso}=\mbox{diag}(E_+,E_-,E_- + E_Z,E_- - E_Z).
\end{equation}
The standard notation is to refer only to the spinor part of the
basis states. The matrix Eq.~(\ref{eq:4x4}) can be rewritten in a more
compact way using the basis of the sixteen sigma matrices,
$\{\sigma_{\alpha,1} \sigma_{\beta,2} \}_{\alpha,\beta=0,x,y,z}$ 
(index 0 denotes a unit matrix; for explicit expressions see App.~\ref{spinmatrices}).
The result is the so-called isotropic exchange Hamiltonian 
(where the constant $E_- - J/4$ was subtracted)
\begin{equation}
H_{iso} = (J/4) \boldsymbol{\sigma}_{1} \cdot \boldsymbol{\sigma}_{2} + \mu {\bf B} \cdot
(\boldsymbol{\sigma}_{1} + \boldsymbol{\sigma}_{2}),
\label{eq:isotropic exchange}
\end{equation}
where the singlet and triplets are separated by the isotropic exchange 
energy $J=E_- - E_+$, the only parameter of the model.

The Hamiltonian Eq.~(\ref{eq:isotropic exchange}) describes the coupling
of the spins in the Heisenberg form. With this form,
the SWAP gate can be performed as the time evolution of the system, 
assuming the exchange coupling $J$ is controllable.
The isotropic exchange has already been studied analytically, in the
Heitler-London, Hund-Mulliken, Hubbard, variational
and other approximations, as well as numerically using
the finite-difference method.
Usually analytical methods provide  a result valid within certain regime of the external
parameters only and a numerical calculation is needed to assess the quality
of various analytical models.

When the spin-orbit interactions are included, additional terms in the
effective Hamiltonian appear, as the matrix elements due to the spin-orbit
interactions $(H_{aniso}')_{ij} = \langle \Phi_i|H_{so}|\Phi_j\rangle$.
Selection rules in Tab.~\ref{table 1} restrict the non-zero matrix elements to
those between a singlet and a triplet,  
\begin{equation}
H_{aniso}'=\left(
\begin{tabular}{cccc}
$ 0 $ & $2 \overline{w}_z$ & $-\sqrt{2}u$&  $\sqrt{2}v$\\
$2 \overline{w}_z^\ast$&$ 0 $&0&0\\ 
$-\sqrt{2}u^\ast$&0&$ 0 $&0\\
$\sqrt{2}v^\ast$&0&0&$ 0 $
\end{tabular}
\right).
\label{4x4}
\end{equation}
Here
$u = (\overline{w}_x+{\rm i} \overline{w}_y),
v= (\overline{w}_x-{\rm i} \overline{w}_y)$
and
\begin{equation}
{\bf \overline{w}}=\langle \Psi_+ | {\bf w}_1 | \Psi_- \rangle, 
\label{matrix element of k}
\end{equation}
where  vector ${\bf w}$ is defined by the spin-orbit interactions
Eq.~(\ref{spinorbit}). 
Using the sigma matrix notation, Eq.~(\ref{4x4}) can be written as 
(see App.~\ref{spinmatrices})
\begin{equation}\label{eq:anisotropic_s}
H_{aniso}'= 
{\bf a'}\cdot (\boldsymbol{\sigma_{1}}-\boldsymbol{\sigma_{2}})+
{\bf b'}\cdot (\boldsymbol{\sigma_{1}}\times \boldsymbol{\sigma_{2}}),
\end{equation}
where the ${\bf a'}$ and ${\bf b'}$ are the spin-orbit vectors defined as
\begin{subequations}\label{eq:spin orbit vectors}
\begin{eqnarray}
{\bf a'} &=& {\rm Re} \langle \Psi_+ | {\bf w}_{1} | \Psi_- \rangle, \\
{\bf b'} &=& {\rm Im} \langle \Psi_+ | {\bf w}_{1} | \Psi_- \rangle.
\end{eqnarray}
\end{subequations}
 The standard exchange Hamiltonian follows as 
\begin{equation}
\label{eq:standard model}
H_{ex}'=H_{iso}+H_{aniso}',
\end{equation}
and we refer to it in further as the first order (effective model) to point the order in which the spin-orbit interactions appear in the matrix elements.
Note that we repeated the derivation of Ref.\onlinecite{Kavokin:PRB2001} additionally including the
external magnetic field.
As we will see below, comparison with numerics shows that treating the spin-orbit interactions to the linear order only is insufficient.

To remedy, we generalize the procedure of Ref.~\onlinecite{Kavokin:PRB2004} to finite magnetic fields. This amounts to repeating the derivation that lead to Eq.~\ref{eq:4x4}, this time starting with the unitarily transformed Hamiltonian Eq.~(\ref{eq:unit_transf}). In this
way, the linear spin-orbit terms are gauged out and the resulting effective Hamiltonian treats
the spin-orbit interactions in the second order in small quantities (the spin-orbit
and the Zeeman couplings).
The transformation asserts that the original Schr\"odinger equation $H_{tot}\Phi=E\Phi$ can be equivalently solved in terms of the transformed quantities $\overline{H}_{tot} (U\Phi) = E (U\Phi)$, with the Hamiltonian $\overline{H}=U H_{tot} U^\dagger$. The transformed Hamiltonian $\overline{H}$ is the same as the original, Eq.~\eqref{ham_tot}, except for the linear spin-orbit interactions, appearing in an effective form $\overline{H}_{so}$.
We again restrict the basis to the lowest four states
and for the spin-orbit contributions we get
\begin{equation}
(H_{aniso})_{ij} = \langle \Phi_i|\overline{H}_{so}|\Phi_j \rangle .
\end{equation}
Using the selection rules and the algebra of the Pauli matrices, we get the exchange Hamiltonian (for obvious reasons, we refer to it as the second order model)
\begin{equation}
\begin{split}
H_{ex}&=(J/4) \boldsymbol{\sigma}_{1} \cdot \boldsymbol{\sigma}_{2} + 
\mu ({\bf B}+{\bf B}_{\rm so}) 
 \cdot (\boldsymbol{\sigma}_{1} + \boldsymbol{\sigma}_{2})\\
& + {\bf a} \cdot (\boldsymbol{\sigma}_{1} - \boldsymbol{\sigma}_{2}) + {\bf b}
\cdot (\boldsymbol{\sigma}_{1} \times \boldsymbol{\sigma}_{2}) - 2K_+.
\end{split} 
\label{eq:final result}
\end{equation}
Compared to the first order model Eq.~\eqref{eq:anisotropic_s}, the functional form of 
the second order model Hamiltonian is the same, except for the effective spin-orbit magnetic field
\begin{equation}
\mu {\bf B}_{so} = {\bf \hat{z}} (K_-/\hbar)  \langle \Psi_- | L_{z,1} | \Psi_- \rangle,
\end{equation}
which appears due to an inversion symmetric part of $\overline{H}_{so}$,
Eq.~(\ref{eq:hsoprime}). The spin-orbit vectors, however, are 
qualitatively different
\begin{subequations}
\label{eq:spin orbit vectors 2}
\begin{eqnarray}
\bf{a} &=& \mu {\bf B} \times {\rm Re} \langle \Psi_+ | {\bf n}_{1} | \Psi_- \rangle,\\
\bf{b} &=& \mu {\bf B} \times {\rm Im} \langle \Psi_+ | {\bf n}_{1} | \Psi_- \rangle.
\end{eqnarray}
\end{subequations}
We remind that the second order effective model Hamiltonian Eq.~(\ref{eq:final result}) refers to the four functions in Eq.~(\ref{eq:basis4}) unitarily transformed
$\{U\Phi_i\}_{i=1,...,4}$. The agreement between the second order effective model and the numerical
data is very good, as we will see below.

\subsection{First order effective Hamiltonian in zero field}

In this section we give $H_{ex}'$ explicitly for zero $B$
and diagonalize it. This is the only case for which is possible
to give an analytical solution.
For zero magnetic field, one can choose the
functions $\Psi_{\pm}$ to be real. Then the matrix elements of
the spin-orbit operator $\bf{w}$ in Eq.~(\ref{spinorbit})
are purely imaginary and ${\bf a}'={\bf 0}$ . 
With the spin quantization axis chosen along the vector ${\bf b}'$, the 4x4 matrix, 
Eq.~\eqref{eq:standard model}, takes the form of
\begin{equation}
H_{ex}'=\left(
\begin{tabular}{cccc}
$-3J/4$ & $2{\rm i}b'$ & 0& 0\\
$-2{\rm i}b'$&$J/4$&0&0\\ 
0&0&$J/4$&0\\
0&0&0&$J/4$
\end{tabular}
\right).
\label{kavokin}
\end{equation}
The upper left $2\times2$ block of this matrix is a Hamiltonian of a spin
$1/2$ particle in a fictitious magnetic field $\boldsymbol{\mathcal{B}}=(0,2b',J/2)/\mu$. 
The eigenstates of this Hamiltonian are spins oriented along the magnetic field $\boldsymbol{\mathcal{B}}$. 
Since the matrix in Eq.~(\ref{kavokin}) is block diagonal, it is easy to see
it can be diagonalized with the help of the following matrix 
\begin{equation}
\Sigma=\left(
\begin{tabular}{cccc}
$0$ & $1$ & $0$ & $0$\\
$1$ & $0$ & $0$ & $0$\\
$0$ & $0$ & $0$ & $0$\\
$0$ & $0$ & $0$ & $0$
\end{tabular}
\right).
\label{pseudo x}
\end{equation}
The Hamiltonian Eq.~\eqref{kavokin} can be diagonalized by $H_{diag} = \varTheta H_{ex}'  \varTheta^\dagger$,
\begin{equation}
H_{diag} = 
\left(
\begin{tabular}{cccc}
$-J/4 -|\mu\mathcal{B}|$ & $0$ & $0$ & $0$\\
$0$&$ -J/4 + |\mu\mathcal{B}|$ & $0$ & $0$\\
$0$ & $0$ & $J/4$ & $0$\\
$0$ & $0$ & $0$ & $J/4$
\end{tabular}
\right),
\label{diagonal}
\end{equation}
where $|\mu\mathcal{B}|^2=4(b')^2+J^2/4$.
In the notation of the Pauli matrices, (see App.~\ref{spinmatrices}),
\begin{equation}
\varTheta=\exp\left(-{\rm i} \frac{\Sigma \theta}{2} \right)=\exp\left(-\frac{{\rm i}}{4} \theta (\sigma_{\mathcal{B},1}-\sigma_{\mathcal{B},2}) \right),
\label{pseudo spin rotation} 
\end{equation}
where $\tan \theta=4b'/J$ and $\sigma_{\mathcal{B}} \equiv \boldsymbol{\sigma}\cdot \boldsymbol{\mathcal{B}} / \mathcal{B}$.

The unitary transformation $\varTheta $ in Eq.~\eqref{pseudo spin rotation} performs the rotation 
of the two spins in the opposite sense. 
The Hamiltonian can be interpreted as a rotation of 
the electron around a spin-orbit field when transferred from one dot to the other.\cite{Kavokin:PRB2001} 
 The spectrum given by Eq.~(\ref{diagonal}) qualitatively differs from
the numerics, which shows there is no influence on the exchange in the second order of the spin-orbit couplings.

\section{Single dot}\label{singledot}

We start with the single dot case, corresponding in our model to $d=0$. The
analytical solution of the single particle Hamiltonian $T+V$ is known
as the Fock-Darwin spectrum.
The corresponding wave functions $\psi$ and the energies 
$\epsilon$ are
\begin{eqnarray}
\label{psiFD}
 \psi_{nl}(r_i,\varphi_i) = C \rho^{|l|}_i e^{-\rho_i^2/2}L_n^{|l|}(\rho_i^2)e^{{\rm i}l\varphi_i},\\
\epsilon_{nl} = \frac{\hbar^2}{m l_B^2} (2n+|l|+1)+B\frac{e\hbar}{2m}l,
\label{energyFD}
\end{eqnarray}
where $\rho_i=r_i/l_B$ and $l_B=[l_0^{-4}+(e B_z /2 \hbar)^2]^{-1/4}$ is the magnetic length; $n$ and $l$ are the radial
and the angular quantum numbers, $C$ is the normalization constant
and $L_n^{|l|}$ are the associated Laguerre polynomials. 

Let us consider now the orbital two electron states $\Psi$,
eigenstates of $H_{orb}$, Eq.~(\ref{ham_orb}).
The Coulomb operator $H_C$ commutes with the rotation 
of both electrons around the $z$ axis, that is, the Coulomb interaction
couples only states with the same total angular momentum. This allows us to label the states with the quantum
number $L=L_1+L_2$, the total angular momentum. Furthermore, the Hamiltonian $H_{orb}$ 
commutes with any spin rotation of any of the electrons, which expresses the fact 
that the Coulomb interaction conserves spin. Therefore we can consider 
the full two electron wavefunctions obtained by supplementing the orbital
part $\Psi$ with a spinor, respecting the overall wavefunction
symmetry, similarly as in Eq.~(\ref{eq:basis4}).

\begin{figure}
\centerline{\psfig{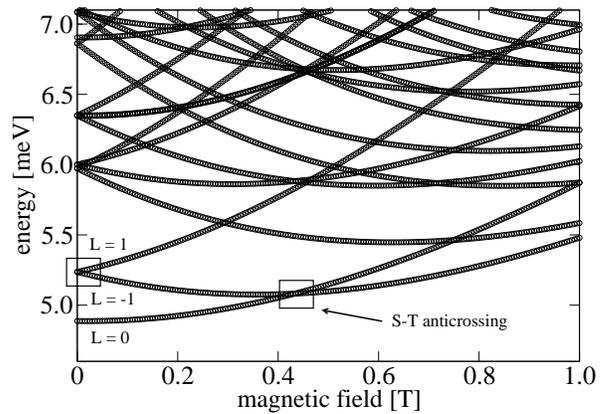}}
\caption{Two-electron energy spectrum of a single dot in perpendicular magnetic field. The 
lowest states are labeled by the total angular momentum $L$. The Zeeman and spin-orbit interactions are neglected. The two regions marked by boxes are magnified on Fig.~\ref{fig:closeB0} and
Fig.~\ref{fig:crossing}.}
\label{fig:2elec_spectrum}
\end{figure}

The two-electron spectrum, without
the Zeeman and the spin-orbit interactions,
is shown in Fig.~\ref{fig:2elec_spectrum}.
At zero magnetic field the ground
state is a non-degenerate singlet state with total angular momentum zero $L=0$. The next two degenerate states are triplets with $L=\pm 1$ and their degeneracy is split by the magnetic field.
Focusing on the two lowest states, most relevant for the qubit pair,
they cross at $B\approx 0.43$ T, so one can turn the ground state from the singlet
to the triplet by applying an external magnetic field.
In the presence of spin-orbit interactions, the crossing is turned 
into anticrossing, as described below.

\subsection{Spin-orbit correction to the energy spectrum in magnetic field}

Suppose some parameter, such as the magnetic field, is being changed.
It may happen at some point that the states of the opposite spin become degenerate.
Such points are called spin hot spots. Here, because of the degeneracy,
 weak spin-orbit interactions have strong effects.
For the spin relaxation, spin hot spots play often a dominant role.\cite{Fabian:PRL1998}

We are interested in the changes to the spectrum due to the spin-orbit
interactions. Let us neglect the cubic Dresselhaus in
this section. To understand the spin-orbit influence,
it is important to note the following commutation relations
for the linear spin-orbit terms 
\begin{equation}
\begin{split}
&[{\bf w}_{BR,1}\cdot \boldsymbol{\sigma}_{1}+{\bf w}_{BR,2}\cdot \boldsymbol{\sigma}_{2},\hat{J}_+] = 0, \\
&[{\bf w}_{D,1}\cdot \boldsymbol{\sigma}_{1}+{\bf w}_{D,2}\cdot \boldsymbol{\sigma}_{2},\hat{J}_-]  = 0,
\end{split}
\end{equation}
where $\hat{J}_{\pm}=\sum_i(\hat{L}_{z,i}\pm \hat{S}_{z,i})$. These commutation rules
hold for any magnetic field $B$. Since the Hamiltonian Eq.~(\ref{ham_orb})
commutes with the operator $\hat{J}_{\pm}$, we can label
the states using the quantum numbers $J_+=L+S_z$
and $J_-=L-S_z$. The spin-orbit interactions
couple only the states with the same quantum numbers $J_+$
and $J_-$, for Bychkov-Rashba and Dresselhaus term, respectively.

\begin{figure}
\centerline{\psfig{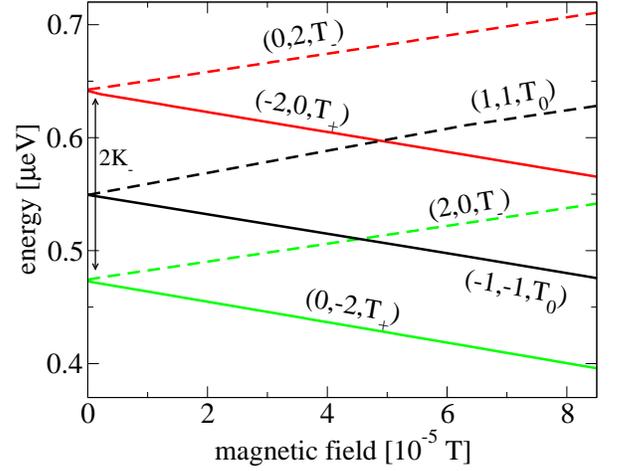}}
\caption{(Color online) Magnified region
from Fig.~\ref{fig:2elec_spectrum}. 
Energy spectrum of a single dot  for small perpendicular magnetic field. Only the states with the total
angular momentum $L=\pm1$ are plotted. A constant shift is removed from the spectrum.
Each state is labeled by the quantum numbers ($J_+,J_-$,$T_i$).}
\label{fig:closeB0}
\end{figure}

Let us focus on the part of the spectrum close to $B=0$ and
on the states with $L=\pm1$, Fig.~\ref{fig:2elec_spectrum}. 
The degeneracy 
of the states is removed by the spin-orbit interactions, as shown in
Fig.~\ref{fig:closeB0}.

Let us now use the Hamiltonian 
Eq.~(\ref{eq:unit_transf}), to understand the influence of the spin-orbit interactions.
The degeneracy of the states with angular momenta $L=\pm1$
makes the description with the lowest two orbital states questionable.
Therefore now we take $3$ orbital states and repeat the derivation
of the second order effective Hamiltonian, obtaining a $7\times7$ matrix.
The basis functions are
\begin{equation}\label{eq:basis7}
\begin{split}
\{ \Phi_i \}_{i=1,...,7} = \{& \Psi_+S, \Psi_{-}T_0, \Psi_{-}T_+, \Psi_{-}T_-, \\
&\Psi_{-}'T_0, \Psi_{-}'T_+, \Psi_{-}'T_-\},
\end{split}
\end{equation}
where $\Psi_{+}$ is the electron wavefunction with
angular momentum $L=0$, and $\Psi_{-}$ and $\Psi_{-}'$ have angular
momentum  $L=+1$, and $L=-1$, respectively. Since the magnetic field is
negligible with respect to the spin-orbit couplings,
the Hamiltonian Eq.~(\ref{eq:hzprime}) is negligible. 
Because of the selection rules, Tab.~\ref{table 1}, 
the contributions 
from Eq.~(\ref{eq:hsoprime}) in the basis Eq.~(\ref{eq:basis7}),
gives non zero matrix elements only for the following pairs,
$\langle\Psi_{-}T_{\pm}|\overline{H}_{so}|\Psi_{-}T_{\pm}\rangle=\pm K_{-}$, and $\langle\Psi_{-}'T_{\pm}|\overline{H}_{so}|\Psi_{-}'T_{\pm}\rangle=\pm K_{-}$.  
For the GaAs parameters, $K_{-}= 0.16\mu$ eV. 
In the region of small magnetic field, 
the states with $J_{\pm} = 0$ are coupled by the spin-orbit interactions
and the lifting is in the second order in the spin-orbit couplings.
The other states are not coupled since they have different values
of $J_{\pm}$. Therefore we conclude that the $2$-orbital state approximation
can be used also for the single dot case (or strongly coupled double dots), 
because the spin-orbit interactions
do not mix the states $\Psi$ and $\Psi^\prime$ in the basis Eq.~(\ref{eq:basis7}). Note that as the coupling is forbidden by the inversion symmetry, the claim holds for an arbitrary oriented magnetic field.

\begin{figure}
\epsfig{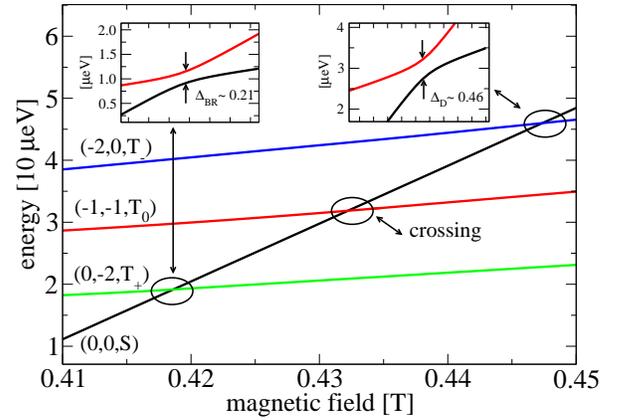}
\caption{(Color online) Lowest energy levels in the anticrossing
region marked in Fig.~\ref{fig:2elec_spectrum}. A constant shift was removed from the spectrum. The quantum numbers ($J_+$,$J_-$,$\Sigma_i$) label the states. Insets  show the anticrossing regions. }
\label{fig:crossing}
\end{figure}

Let us now discuss the second degeneracy region marked in Fig.~\ref{fig:2elec_spectrum}, magnified in Fig.~\ref{fig:crossing}.
The spin-orbit interactions induce two anticrossings.
The first is due to the Bychkov-Rashba term, since the crossing states
have different $J_-$, but the same $J_+=0$ and couples the singlet $S$
and triplet $T_+$.
The second is due to the Dresselhaus term which
couples states with $J_-=0$, the singlet $S$ and the triplet $T_-$.
The central point is a crossing point, because 
the crossing state differ in both $J_+$ and $J_-$.
The splitting energy can be evaluated using the unitarily transformed
Hamiltonian Eq.~(\ref{eq:unit_transf}). Using the degenerate
perturbation theory, one can estimate analytically, using
Eq.~(\ref{eq:hsoprime}) and Eq.~(\ref{eq:hzprime}), the value of
the two gaps to be
$\Delta_{BR}  \approx 4\sqrt{2}\mu B l_0/l_{BR}=0.15\mu eV$ and 
$\Delta_{D}  \approx 4\sqrt{2}\mu B l_0/l_{D}=0.58\mu eV$.
These values are consistent
with the numerical values. 

\section{Double dot}\label{doubledot}
The double dot denotes the case when the interdot distance is of the order of the confinement length.
In the next sections
we discuss our effective models, Eq.~(\ref{eq:final result})
and Eq.~(\ref{eq:standard model}) in the double dot regime and compare them with numerics.

\subsection{Heitler-London approximation}
The analytical solution for the two electron wavefunctions in a double dot potential is not known.
We consider here the Heitler-London ansatz since
it is a good approximation at large interdot distances and we can
work out the spin-orbit influence on the spectrum analytically. For this purpose,
we compute the spin-orbit vectors, Eq.~(\ref{eq:spin orbit vectors})
and Eq.~(\ref{eq:spin orbit vectors 2}),
for our models.
 
In the Heitler-London ansatz,
the two electron eigenfunctions are given  by 
\begin{equation}
\Psi_{\pm}  = \frac{1}{\sqrt{2(1\pm |\langle \psi_{L,1} | \psi_{R,1}\rangle| ^2)}} 
(|\psi_{L,1}\rangle |\psi_{R,2}\rangle \pm |\psi_{R,1}\rangle |\psi_{L,2}\rangle),
\label{states pm}
\end{equation}
where $|\psi_{L(R),i}\rangle$ is a single electron Fock-Darwin state centered
in the left (right) dot occupied by the $i$-th electron.
Below, in Eqs.~\eqref{spin orbit vectors lr}-\eqref{eq:aux}, we skip the particle subscript $i$, as the expressions contain only single particle matrix elements (all $\psi$, $w$, $n$, $L_z$ would have the same subscript, say $i=1$).
With this ansatz,
the spin-orbit vectors, Eq.~\eqref{eq:spin orbit vectors},
follow as
\begin{subequations}\label{spin orbit vectors lr}
\begin{eqnarray}
{\bf a'} & =& \frac{1}{\sqrt{1-|\langle \psi_L | \psi_R \rangle |^4}} 
\langle \psi_L | {\bf w} | \psi_L \rangle,\\
{\bf b'} & = &\frac{{\rm i}}{\sqrt{1-|\langle \psi_L | \psi_R \rangle |^4}} 
\langle \psi_L | {\bf w} | \psi_R \rangle \langle \psi_R | \psi_L \rangle.
\end{eqnarray}
\end{subequations}

Similarly we get the spin-orbit vectors, Eq.~(\ref{eq:spin orbit vectors 2}), 
as
\begin{subequations}\label{so vectors standard}
\begin{eqnarray}
{\bf a} &=& \frac{\mu}{\sqrt{1-|\langle \psi_L | \psi_R \rangle |^4}} 
\langle \psi_L 
| {\bf B} \times{\bf n}|\psi_L\rangle,\\
{\bf b} &=& \frac{{\rm i\mu}}{\sqrt{1-|\langle \psi_L | \psi_R \rangle |^4}} \langle \psi_L
|{\bf B} \times{\bf n}|\psi_R\rangle \langle \psi_R | \psi_L \rangle,
\end{eqnarray}
\end{subequations}
and the spin-orbit induced magnetic field 
\begin{equation}
\begin{split}
\mu {\bf B}_{\rm so} = {\bf \hat{z}} \frac{K_-/\hbar}{1-|\langle \psi_L | \psi_R \rangle |^2} 
\Big(\langle \psi_L | L_z | \psi_L\rangle + \\ 
-\langle \psi_L | L_z | \psi_R\rangle 
\langle \psi_R | \psi_L \rangle\Big).
\end{split}
\label{eq:aux}
\end{equation}
The explicit formulas for the vectors in Eqs.~(\ref{spin orbit vectors lr})-(\ref{eq:aux}) are in App.~\ref{heitlerlondon}.
Differently from the spin-orbit vectors in Eq.~\eqref{spin orbit vectors lr}, 
the vectors in Eq.~\eqref{so vectors standard} reveal explicitly 
the anisotropy with respect to the magnetic field  and dot orientation\cite{OlendskiPRB:2007, PRL-104-246801} 
(note that $x$ and $y$ in the definition of $\bf{n}$, Eq.~(\ref{direction}) are the crystallographic coordinates).

\subsection{Spin-orbit correction to the energy spectrum in zero magnetic field}

In the previous sections, we have derived two effective Hamiltonians,
$H_{ex}'$, and $H_{ex}$, given by Eqs.~\eqref{eq:final result}-\eqref{eq:spin orbit vectors 2} and Eqs.\eqref{eq:anisotropic_s}-\eqref{eq:standard model}, respectively. We now compare the energy spectrum given by these models with exact numerics. We present the spin-orbit induced energy shift, the difference between a state energy if the spin-orbit interactions are considered and artificially set to zero. For each model we examine also its Heitler-London approximation, which yields analytical expressions for the spin-orbit vectors, as well as the isotropic exchange energy (given in Sec.~IV.A and Appendix C). Thus, the effective models in the Heitler-London approximation (we denote them by superscript HL) are fully analytic. The non-simplified models (we refer to them as ``numerical'') require the two lowest exact double dot two-electron wavefunctions, which we take as numerical eigenstates of $H_{orb}$.

Apart from the energies, we compare also the spin-orbit vectors. 
Since they are defined
up to the relative phase of states $\Psi_+$ and $\Psi_-$, the observable
quantity is $c'=\surd (a')^2+ (b')^2$ and analogously for $c=\surd a^2+ b^2$.
We refer to these quantities as the anisotropic part of the exchange coupling.

\begin{figure}
\centerline{\psfig{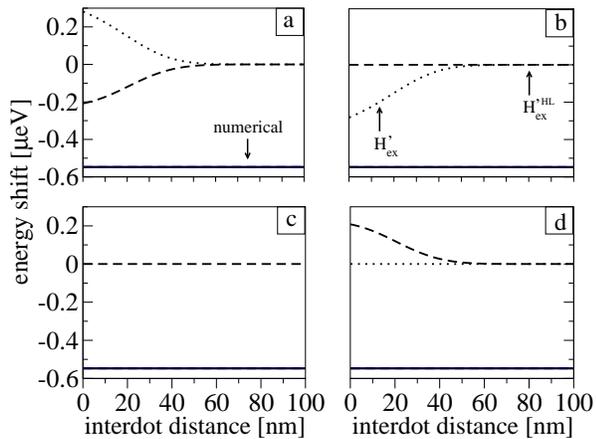}}
\caption{Spin-orbit induced energy shifts at zero 
magnetic field as a function of the interdot distance. Exact numerics (solid), first order model $H_{ex}'$ (dotted) and first order model in HL approximation (dashed) are given.
a) Singlet, b) triplet $T_0$, c)  triplet $T_+$, d) triplet $T_-$.
The results of the second order model (both $H_{ex}^{HL}$ and $H_{ex}$ give the same) are indiscernible
from exact numerical data.}
\label{fig:socont_B0}
\end{figure}

Figure~\ref{fig:socont_B0} shows the spin-orbit induced energy shift
as a function of the interdot distance for each of the four states.
The  exact numerics gives a constant and equal shift for all $4$ spin states,
with value $-0.54\mu$eV. Let us consider the second order model, Eq.~(\ref{eq:final result}).
For zero magnetic field, all spin-orbit vectors are zero, as is the
effective magnetic field. 
The only contribution comes from the constant
term $2K_+ = -0.54\mu$eV that is the same for all states. Our derived spin-model,
Eq.~(\ref{eq:final result}),
accurately predicts the spin-orbit contributions to the energy.
On the other hand, the first order models  $H_{ex}'^{HL}$ and $H_{ex}'$ are completely off 
on the scale of the spin-orbit contributions. 
The exchange Hamiltonian $H_{ex}'$ does not predict the realistic 
spin-orbit influence on the spectrum, even in the
simple case when the magnetic field is zero.

\begin{figure}
\centerline{\psfig{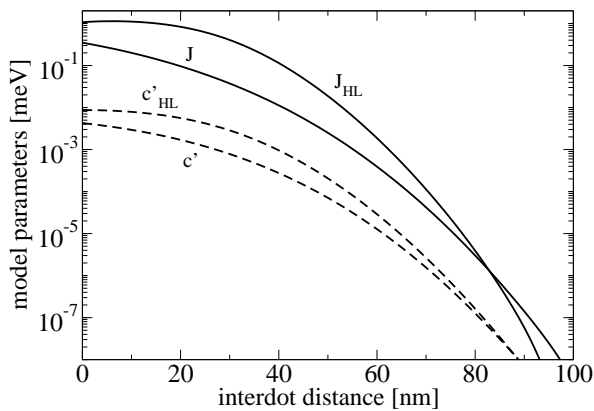}}
\caption{Spin-orbit parameters at zero magnetic field as function of the interdot distance.
Numerical value and the Heitler-London approximation for the isotropic exchange (solid) and the anisotropic exchange of the first order model (dashed).}
\label{fig:param_B0}
\end{figure}

Figure~\ref{fig:param_B0} shows the non zero parameters for all
four models.
The exact isotropic exchange $J$ decays exponentially with the interdot distance. 
The same behavior is predicted in the Heitler-London approximation.
It decays exponentially, but deviates
from the numerical results. As for the anisotropic exchange,
the first order model $H_{ex}'$ gives an exponentially falling spin-orbit parameter $c'$,
an order of magnitude smaller than $J$. In contrast, the second order model $H_{ex}$ 
predicts zero spin-orbit anisotropic  exchange.
First main result, proved numerically and
justified analytically by the Hamiltonian $H_{ex}$, is that
{\it at zero magnetic field the 
spin-orbit vectors vanish},  up to the second order in
spin-orbit couplings at any interdot distance.  
In the transformed basis, there is no anisotropic exchange at the
zero magnetic field due the spin-orbit interactions, 
an important result for the quantum computation.
Indeed, since the exchange energy can be used 
to perform a SWAP operation, this means that the spin-orbit interactions
do not induce any significant errors on the gate operation. The only difference is the computational basis,
which is unitarily transformed with respect to the usual singlet-triplet basis. 

\subsection{Finite magnetic field}

\begin{figure}
\centerline{\psfig{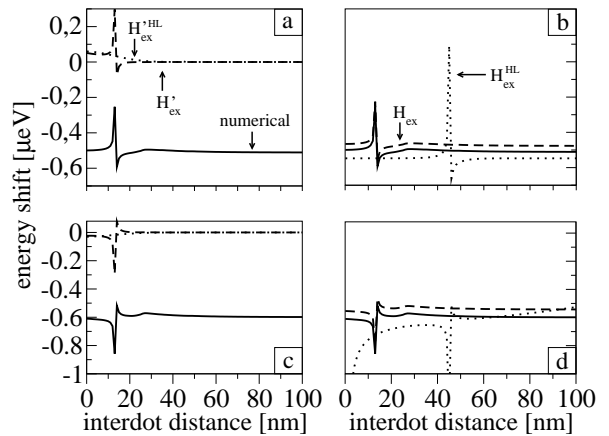}}
\caption{Spin-orbit  induced energy shifts at $1$ Tesla perpendicular magnetic field versus
the interdot distance.
a) Energy shift of the Singlet $S$ in the exact numerics (solid) is compared to the numerical (dashed) and the Heitler-London approximation (dotted) first order model. In b) similar comparison is made for the second order model. Panels c-d) are analog of a-b) showing the energy shifts of the triplet $T_{+}$.
}
\label{fig:socont_B1}
\end{figure}

In the presence of a perpendicular magnetic field the structure
of the spin-orbit contributions are quite different with respect to the
zero field case. First of all,
anticrossing points appear, where the energy shift is enhanced.
Figure~\ref{fig:socont_B1} shows the spin-orbit contributions
 in a finite magnetic field . We plot only the anticrossing states, the singlet $S$ and
the triplet $T_{+}$.
The prediction of the first order model is shown in the left panels of 
Fig.~\ref{fig:socont_B1}. 
As in the case of zero magnetic field, this model is off from the
numerical results. In particular, it still predicts a zero contribution, except close to the anticrossing point.
We note that the discrepancy is not connected to (a failure of) the Heitler-London
approximation, as using the exact numerical two electron wavefunctions does not
improve the model predictions.

In the right panels of Fig.~\ref{fig:socont_B1},  the comparison between
the second order model and the numerics is provided. We observe
that the  model is very close to the numerics, even  though the Heitler-London approximation predicts
the crossing point in a different position. The predictions
of the numerical second order model $H_{ex}$ is consistent with the exact numerics. The only discrepancy
is due to the influence of the cubic Dresselhaus term, as we will
see in the next section.
 
\begin{figure}
\centerline{\psfig{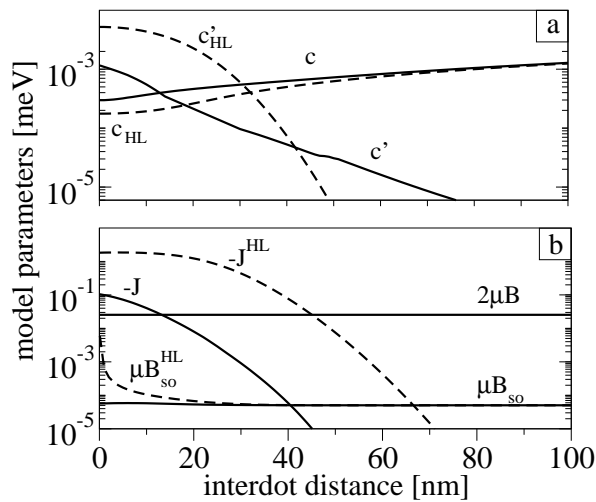}}
\caption{Spin-orbit parameters at $1$ Tesla perpendicular magnetic field versus the interdot distance.
a) Numerical (solid) and Heitler-London approximation (dashed) anisotropic exchange vectors for the first and second order model. b) Isotropic exchange, Zeeman energy and the spin-orbit induced effective  magnetic field.}
\label{fig:param_B1}
\end{figure}

To get more insight, in Fig.~\ref{fig:param_B1}
we have plotted the parameters of the models.
Fig.~\ref{fig:param_B1}a  shows the anisotropic  exchange strengths
in the two models. The first order model $H_{ex}'$  predicts the anisotropic exchange decreasing
with the interdot distance, similar to the isotropic exchange energy.
For large interdot distance the anisotropic exchange $c'$ disappears. 
This means
there is no influence on the energy due to the spin-orbit interactions. 
On the other hand, for the second order model $H_{ex}$ the conclusion is different.
For large interdot distances $c^{HL}$ and $c$ are linear in $d$.
Furthermore, the anisotropic exchange  computed in the
Heitler-London ansatz is very close to the numerical one.
We make a very important observation here: surprisingly, concerning the anisotropic exchange the Heitler-London
is quite a good approximation for all interdot distances even in a finite magnetic field. Therefore, despite
its known deficiencies to evaluate the isotropic exchange $J$, it grasps
the anisotropic exchange even quantitatively, rendering the spin-orbit
part of the second order effective Hamiltonian $H_{ex}$ fully analytically. One can understand this looking at Eqs.~(\ref{eq:spin orbit vectors 2}). The anisotropic exchange vectors are given by the dipole moment of the matrix element between the left and right localized state (see App. C for explicit formula). This dipole moment is predominantly given by the two local maxima of the charge distribution (the two dots) and is not sensitive to the interdot barrier details, nor on the approximation used to estimate the lowest two orbital two-electron states. This is in strong contrast to the isotropic exchange, which, due to its exponential character, depends crucially on the interdot barrier and the used approximation.

Figure~\ref{fig:param_B1}b shows the isotropic exchange $J$, and the
effective magnetic field induced by the spin-orbit interactions $\mu B_{so}$
compared to the Zeeman energy $2\mu B$. 
We see the failure
of the Heitler-London approximation for $J$. Although the numerical calculation and
the analytical prediction have the same sign (this means that the ground state
is the triplet) they differ by an order of magnitude. The Zeeman energy
is constant and always much larger than the effective spin-orbit
induced magnetic field $\mu B_{so}$. 
Consequently, the effective field can be always neglected.
The point where the Zeeman energy equals to the isotropic
exchange (close to $d=18$nm) is the anticrossing point, where the spin-orbit
contributions are strongly enhanced, as one can see in Fig.~\ref{fig:socont_B1}.

\begin{figure}
\centerline{\psfig{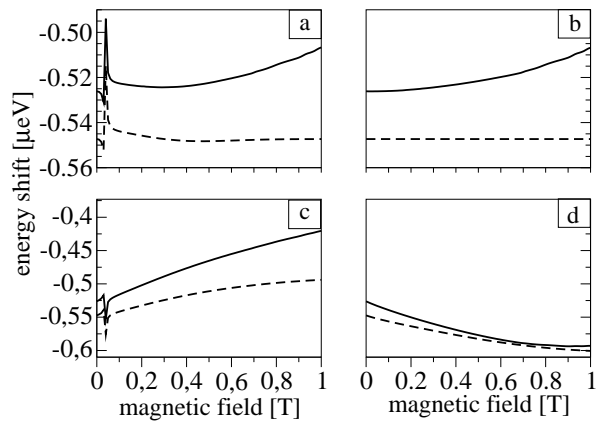}}
\caption{Spin-orbit induced energy shifts of a double dot system with
interdot distance of $55$ nm versus the perpendicular magnetic field.
 a) singlet S, b) triplet $T_{0}$, c) triplet $T_{+}$,
d) triplet $T_{-}$. Exact numerics (solid) and the numerical second order model $H_{ex}$ (dashed).}
\label{fig:socont_d55nm}
\end{figure}

Let us consider a double
dot system at fixed interdot distance of $55$nm, corresponding to zero field
isotropic exchange of $1\mu$eV. In Fig.~\ref{fig:socont_d55nm}
the spin-orbit contributions versus the magnetic field are plotted for the second order model $H_{ex}$ 
and the exact numerics. 
We can conclude that to describe the spin-orbit influence on the
states in a double-dot system it is important to use the second order Hamiltonian $H_{ex}$. 

\begin{figure}
\centerline{\psfig{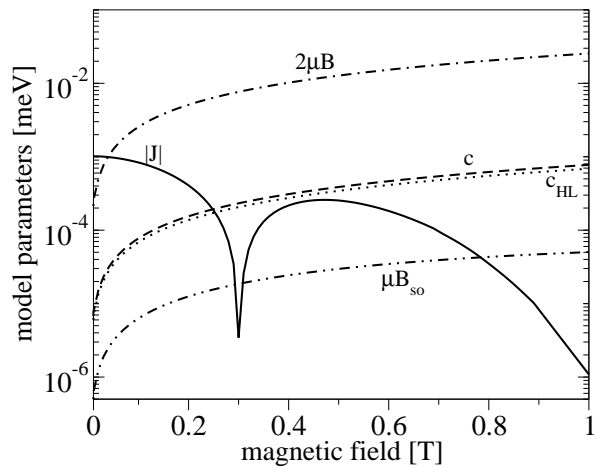}}
\caption{Spin-orbit parameters of the second order numerical model $H_{ex}$ in a double dot system with
interdot distance of $55$ nm versus the magnetic field.}
\label{fig:param_d55nm}
\end{figure}

In Fig.~\ref{fig:param_d55nm}
the spin-orbit parameters versus the magnetic field are plotted.
The
main influence on the spin is due to the Zeeman interaction in the whole range of $B$,
since $\mu B_{so}$ is several orders of magnitude smaller than the Zeeman energy.
At the ground state anticrossing point, the isotropic exchange crosses
zero, while the anisotropic parameter $c$ is finite, leading to spin hot spots. Apart from these, since
the anisotropic exchange is two orders of magnitude smaller than the 
Zeeman energy, the spin-orbit induced
energy shifts are minute. 

\subsection{Cubic Dresselhaus contributions}\label{cubicdresselhaus}

\begin{figure}
\centerline{\psfig{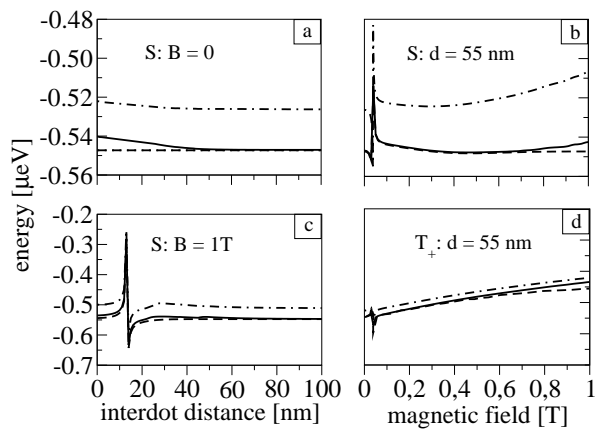}}
\caption{The spin-orbit induced energy shift as a function of the interdot distance (left panels) 
and perpendicular magnetic field (right panels). a) Singlet 
in zero magnetic field, c) singlet at 1 Tesla field, b) and d) singlet and
triplet $T_+$ at $55$ nm. The numerical second order  model $H_{ex}$ (dashed line), exact numerics 
(dot-dashed line) and exact numerics without the cubic Dresselhaus  term (solid line)
.}
\label{fig:cubic}
\end{figure}

Finally we consider the role of the cubic Dresselhaus term.
The Schrieffer-Wolff transformation does not remove it in the linear order. 
Figure~\ref{fig:cubic} shows the energy shifts induced by the spin-orbit
interactions also in the case where we do not take into account
the cubic Dresselhaus term.
One can see a very good agreement
between  the second order model $H_{ex}$ and the exact numerics where the cubic Dresselhaus term was omitted.
Therefore we can conclude that the main part of the discrepancy
we see in the spin-orbit induced energy shifts are due
to the cubic Dresselhaus term. 

\section{Conclusions}
We analyzed the spin-orbit influence on two electrons confined in a lateral double quantum dot. We focused on the lowest part of the Hilbert space, which corresponds to a qubit pair. In Ref.~\onlinecite{Baruffa:PRL2010} a Hamiltonian for such pair was proposed, with the spin-orbit interactions giving rise to an anisotropic exchange interaction. Within a unitarily transformed basis, this interaction is encoded into two real three dimensional spin-orbit vectors. These, together with the isotropic exchange energy and the magnetic field vector, completely parametrize an effective two qubit Hamiltonian. In this work, we examined the quantitative validity of this effective Hamiltonian.

In addition to a numerical study, we also provided the details of the effective Hamiltonian derivation, which were skipped in Ref.~\onlinecite{Baruffa:PRL2010}. We noted that it can be diagonalized analytically if the effective spin-orbit vectors are all aligned with the external magnetic field, the only exactly solvable case (apart from the trivial case of no spin-orbit interactions present). We also evaluated the spin-orbit vectors in the Heitler-London approximation and compared the analytical results with their exact numerical counterparts.

There are three possible sources for a discrepancy between the model and the exact data: the higher excited orbital states of the quantum dot, the higher orders of the effective (unitary transformed) spin-orbit interactions and the cubic Dresselhaus term. Elucidation of their importance is one of the main results of this work. i) We find the cubic Dresselhaus term is the main source of the discrepancy. In a typical double dot regime and a moderate field of 1 Tesla, it brings an error of $\sim 0.1\mu$eV for the energies, while the two other mentioned corrections have an order of magnitude smaller influence. ii) We find the effective Hamiltonian describes both the weak and the strong coupling regimes (the single dot represents the strongest possible coupling). iii) Surprisingly, the spin-orbit vectors obtained within the Heitler-London approximation are faithful even at a finite magnetic field. Overall, we find the anisotropic exchange Hamiltonian to be generally reliable, providing a realistic and yet simple description for an interacting pair of spin qubits realized by two coupled quantum dots.

\acknowledgments
We would like to thank Guido Burkard for useful discussion, Martin Gmitra
and Andrea Nobile for numerical advice. 
This work was supported by DFG GRK 638, SPP 1285, SFB 689,  
NSF grant DMR-0706319, RPEU-0014-06, ERDF OP R\&D Project ``QUTE'', 
CE SAS QUTE and DAAD.

\appendix
\section{Numerical method}\label{appendix}
Here we describe the numerical method we use to diagonalize the two electron Hamiltonian Eq.~(\ref{ham_tot}).
We proceed in three steps.\cite{Tureci:PRB2006}
We first diagonalize the single electron Hamiltonian $H = T + V$, using
the numerical finite differences method with the Dirichlet boundary condition
(vanishing of the wave function at boundaries). 
We do not consider the spin dependent part (spin-orbit, Zeeman) at this step. 
This allows us to exploit the symmetries of the confinement potential.
The single electron Hamiltonian is diagonalized by the Lanczos method\cite{Stano:PRB2005}.
The typical number of points in the grid we use
is $60\times 60$, giving relative precision of the energy of order $10^{-6}$.

In the second step, using
the obtained single electron eigenstates $\{(\psi_i,\epsilon_i)\}$,
we construct the two electron states. We use them as a basis in which
the two electron orbital Hamiltonian Eq.~(\ref{ham_orb}) is diagonalized.
The two-electron states are constructed as symmetric
\begin{eqnarray}\label{symm_fun}
|\Psi_s^{(i,j)}\rangle &=& \frac{1}{\sqrt{2}}
(|\psi_{i,1}\rangle|\psi_{j,2}\rangle + |\psi_{j,1}\rangle|\psi_{i,2}\rangle ) \mbox{     for } i\neq j,\\
|\Psi_s^{(i,j)}\rangle &=& |\psi_{i,1}\rangle|\psi_{j,2}\rangle \mbox{     for } i= j,
\end{eqnarray}
and antisymmetric 
\begin{eqnarray}\label{antisymm_fun}
|\Psi_t^{(i,j)}\rangle = \frac{1}{\sqrt{2}}
(|\psi_{i,1}\rangle|\psi_{j,2}\rangle - |\psi_{j,1}\rangle|\psi_{i,2}\rangle ),
\end{eqnarray}
with respect to the particle exchange. We choose $n_{s.e.}$ 
single electron orbitals, typically $n_{s.e.}=21$. 
The total number of the two particle states is then $n_{s.e.}^2$.

The spatial symmetry allows us to reduce
the dimension of the two
electron Hamiltonian matrix to diagonalize.
Namely, the matrix is block diagonal, with the basis functions
grouped according to the spatial symmetry ($1$, $x$,$y$, $xy$) and
particle exchange symmetry ($\pm1$).
This results in $8$ blocks and holds for zero perpendicular magnetic field.
In a finite field, we get $4$ blocks, as there are only two
spatial symmetries possible ($1$, and $x$). Each block is diagonalized
separately.

The matrix element of the two-electron Hamiltonian, Eq.~(\ref{ham_orb}),
in our basis is
\begin{equation}
\begin{split}
&\langle\Psi_a^{(i,j)}|H_{orb}
|\Psi_b^{(n,m)}\rangle 
=(\epsilon_i+\epsilon_j)\delta_{i,m}\delta_{j,n}\delta_{a,b}+\\
+&\delta_{a,b}\int d\mathbf{r}_1 \int d \mathbf{r}_2 
\Psi_a^{(i,j)} \frac{e^2}{4\pi\varepsilon_0\varepsilon_r}
\frac{1}{| \mathbf{r}_1 - \mathbf{r}_2 |}\Psi_b^{(n,m)}.
\end{split}
\label{coulomb_int}
\end{equation}
The last term in Eq.~(\ref{coulomb_int}) is due to the Coulomb interaction
and it leads to off diagonal terms in the Hamiltonian.
We diagonalize the matrix defined in Eq.~(\ref{coulomb_int})
to get the eigenspectrum $\{(\Psi_i,E_i)\}$.

In the third step, we add the spin dependent parts to the Hamiltonian.
We construct a new basis by expanding the wavefunctions
obtained in the previous step by the spin.
The orbital wavefunction $\Psi_{i}$ gets 
the spinor according to its particle exchange symmetry. 
The symmetric function gets the singlet  $S$ while the antisymmetric
appears in three copies, each with one of the three triplets 
$T_0$, $T_{+}$ and $T_-$. We denote the new states by
\begin{equation}\label{eq:2el_states}
|\Phi_{i\Sigma}\rangle = |\Psi_i\rangle|\Sigma \rangle,
\end{equation}
where $|\Sigma \rangle$  corresponds
to one of the 4 spin states.  The matrix elements of the
total Hamiltonian Eq. (\ref{ham_tot}) are
\begin{equation}
\begin{split}
\langle \Phi_{i\Sigma}|H_{tot}|\Phi_{i'\Sigma'}\rangle &= 
E_i\delta_{i,i}\delta_{\Sigma,\Sigma'}  + \\
&+ 2\mu |{\bf B}|
(\delta_{\Sigma,T_{+}}-\delta_{\Sigma,T_{-}}))\delta_{i,i}\delta_{\Sigma,\Sigma'}+ \\
&+ \sum_{j=1,2}\langle \Psi_i|{\bf w}_j|\Psi_{i'}\rangle \cdot \langle \Sigma|\boldsymbol \sigma_j |\Sigma'\rangle ,
\end{split}
\end{equation}
where the last term is the matrix element of the spin-orbit interactions.
The resulting matrix is diagonalized to get the final eigenstates. 
We choose a certain number $n_s$
of lowest $\Psi_i$ states, depending on the required precision.
 In our simulations $n_s = 250 $, resulting to the accuracy
of the order of $10^{-5}$ meV for the energy.

\subsubsection*{Coulomb integral}
Computationally most demanding are the Coulomb integrals. 
Indeed, the typical size of the Hamiltonian matrix, in the second step, 
is $441\times441$, requiring 
at least $10^6$ Coulomb integrals. Writing 
functions involved in the Eq. (\ref{coulomb_int}) as Slater determinants, we can express
the integral as a sum of terms such as the following 
\begin{equation}
 \begin{split}
C_{ijkl} &= \frac{e^2}{4\pi\varepsilon_0\varepsilon_r}
\int d\mathbf{r}_1  d \mathbf{r}_2 
\frac{ \psi_i(\mathbf{r}_1)^* \psi_j(\mathbf{r}_2)^* \psi_k(\mathbf{r}_1) \psi_l(\mathbf{r}_2)}
{| \mathbf{r}_1 - \mathbf{r}_2 |}=\\
&= \frac{e^2}{4\pi\varepsilon_0\varepsilon_r} \int d\mathbf{r}_1  d \mathbf{r}_2 \frac{\mathcal{F}_{ik}(\mathbf{r}_1) \mathcal{F}_{jl}(\mathbf{r}_2)}
{| \mathbf{r}_1 - \mathbf{r}_2 |},
\end{split}
\end{equation}
where $\mathcal{F}_{ik}(\mathbf{r})= \psi_i(\mathbf{r})^* \psi_k(\mathbf{r})$.
The symmetry of the Coulomb integral $C_{ijkl}=C_{jilk}$ reduces the number of needed
matrix elements to a half. For the single dot, $\psi_i$ are the Fock-Darwin
functions and it is possible to derive an analytical formula for $C_{ijkl}$.
In our case, since the single particle functions are given numerically,
we have performed a numerical integration.
Using the Fourier transform, we can reduce the $4$-dimensional integration
to two dimensional
\begin{equation}\label{fourier_integral}
 C_{ijkl}= 2\pi\frac{e^2}{4\pi\epsilon_0\epsilon_r}\int d\mathbf{q} 
\tilde{\mathcal{F}}_{ik}(\mathbf{q} ) \tilde{\mathcal{F}}_{jl}(-\mathbf{q} )\frac{1}{|\mathbf{q}|},
\end{equation}
where
\begin{equation}
\tilde{\mathcal{F}}_{ik}(\mathbf{q})=\frac{1}{2\pi}\int d\mathbf{r}
\mathcal{F}_{ik}(\mathbf{r})\exp({\rm i} \bf{q}\cdot\bf{r}).
\end{equation}
For the evaluation of the
Fourier transforms, we use the Discrete Fourier Transform algorithm with
the attenuation factors, as described in Ref.\onlinecite{NRecipes:2007}.

We compute (\ref{fourier_integral}) according to the
perturbative formula
\begin{equation}
\begin{split}
&C_{ijkl} = \sum_{n,m}^{N_x,N_y}\sum_{k_1,k_2=0}^{k_1+k_2\leq N}\sum_{l_1,l_2=0}^{k_1,k_2}I(l_1,l_2,n,m)\times\\
&\times\frac{(-q_n)^{(k_1-l_1)}}{(k_1-l_1)!l_1!}
\frac{(-q_m)^{(k_2-l_2)}}{(k_2-l_2)!l_2!}\partial_x^{k_1}\partial_y^{k_2}
f({\bf q})\vert_{q_{nm}},
\end{split}
\end{equation}
where $f({\bf q})\vert_{q_{nm}}=\tilde{\mathcal{F}}_{ik}(\mathbf{q} ) \tilde{\mathcal{F}}_{jl}(-\mathbf{q} )$ is calculated in the point $q_{nm}$,
$N$ is the perturbative order (the order of the Taylor expansion),
$N_x$ and $N_y$ are the number of grid points in the $x$ and in the $y$ direction,
respectively. The coefficients $I(l_1,l_2,n,m)$ depend only on the geometry
of the grid and are defined as
\begin{equation}\label{integral_I}
I(l_1,l_2,n,m) = \int_{\Omega_{x}} dx \int_{\Omega_{y}} dy \frac{x^{l_1}y^{l_2}}{\sqrt{x^2 + y^2}}.
\end{equation}
Here $\Omega_{x}=\left\langle(n-1/2)\delta_x,(n+1/2)\delta_x\right\rangle$ is the 
integration region 
and $\delta_x$ is the grid spacing along $x$. Similarly for the $y$ direction.
In our simulations we use the previous formula up to the 4-nd order in the Taylor expansion.
The achieved relative precision is $10^{-5}$, with the computational time for one
Coulomb element $\approx 50$ ms.

\section{Two electron symmetry}\label{SDsymmetries}

Suppose
the single particle Hamiltonian commutes with certain set of operators
$\{O_{\alpha} \}$, and therefore the single particle states $\psi_i$
can be chosen such that they
have definite symmetries forming a representation of the group $O$ of the
symmetry operators
\begin{equation}
 O_{\alpha}\psi_i = o_{\alpha}^{i} \psi_i.
\end{equation}
For example, since the double dot potential has inversion symmetry along $x$ axis,
$I_x$ is in the group $O$, while $o_{x}^{i} = \pm 1$ -- the states are
symmetric or antisymmetric with respect to $x$ inversion. Now consider the
two electron states $|\Psi_{s/t}^{(i,j)}\rangle$,
Eq.~(\ref{symm_fun}-\ref{antisymm_fun}). 
These states also have definite symmetry if a certain
operator from $O$ acts simultaneously on both particles
\begin{equation}
O_{\alpha,1} O_{\alpha,2}\Psi_{ij} = 
o_{\alpha}^{i}o_{\alpha}^{j} \Psi_{ij}.
\end{equation}
For our case of
the symmetry group $C_{2v}$, since $o_{\alpha}^{i} = \pm 1$, the set of all possible
products of two characters is the same as the set of characters for
a single particle,
$\{o_{\alpha}^{i}o_{\alpha}^{j}\}_{i,j}=
\{o_{\alpha}^{i}\}_{i}$. This means the
two particle states will form the same symmetry classes as single particle states
with the same characters. 

\section{Heitler-London approximation}\label{heitlerlondon}
In the Heitler-London approximation, the exchange energy is calculated as
\begin{equation}
 J_{HL} = \langle \Psi_- |H_{orb}| \Psi_- \rangle - \langle \Psi_+ |H_{orb}| \Psi_+ \rangle
\end{equation}
with the functions $\Psi_{\pm}$ given in Eq.~(\ref{states pm}). 
The single particle ground state wavefunction of the Fock-Darwin
spectrum is
\begin{equation}
\psi_{00}(x,y)=\frac{1}{l_B\sqrt{\pi}}\exp\left[{-\frac{x^2+y^2}{2l_B^2}}\right],
\end{equation}
where $l_B$ is the effective confinement length defined by 
$l_B^2=l_0^2/\sqrt{1+B^2e^2l_0^4/4\hbar^2}$.
The wavefunctions $\psi_{L(R)}$ are obtained shifting the 
Fock-Darwin ground state to $(\pm l_0d,0)$. 
In the presence of the magnetic field we have to
add a phase factor because of the gauge transformation
$\vec{A}'=B/2(-y,x\pm d)\rightarrow \vec{A}=B/2(-y,x) $; we have
\begin{equation}
\begin{split}
\psi_{L(R)}= \exp{\left[\pm {\rm i} d\zeta\vartheta \frac{y}{l_0}\right]}\psi_{00}(x\pm l_0d,y),\\
\zeta=\left(\frac{l_0}{l_B}\right)^2,\;\;\; \vartheta=\frac{Bel_B^2}{2\hbar},
\;\;\;l_B=l_0(1-\vartheta^2)^{1/4}.
\end{split}
\end{equation}
The overlap between the left and right functions is
\begin{equation}
\Omega = \langle \psi_L|\psi_R\rangle = \exp{\left[-\zeta d^2\right(1+\vartheta^2)]},
\end{equation}
and the exchange energy is
\begin{equation}\label{eq:iso-exchange}
\begin{split}
J_{HL}&=\frac{\hbar\omega_0}{\sinh[2\zeta d^2(1+\vartheta^2)]}
\Big\{c_s\sqrt{\zeta}\big(\exp{[-\zeta d^2]}I_0(\zeta d^2)+\\
&-\exp{[\zeta d^2\vartheta^2]} I_0(\zeta d^2\vartheta^2)\big)+
\frac{2d}{\sqrt{\pi \zeta}}\left(
1-\exp{[-\zeta d^2]}\right)+ \\
&+ 2d^2 \left( 1-\mbox{Erf}(d\sqrt{\zeta}) \right) \Big\},
\end{split}
\end{equation}
where $I_0$ is the zeroth-order modified Bessel function of the first kind.
The factor $c_s$ is the  ratio between the Coulomb strength and the confinement
energy,
$c_s=e^2\sqrt{\pi/2}/4\pi\varepsilon_0\varepsilon_r l_0\hbar\omega_0$.
Similar formula can be found in Ref.\onlinecite{Burkard:PRB1999} for a quartic confinement
potential. The formula~(\ref{eq:iso-exchange}) has been derived in
Ref.\onlinecite{Pedersen:PRB2007} (in the original paper there is a trivial typo that we correct).

The two electron energies for the states $\Psi_-$ and $\Psi_+$ are
\begin{equation}
E_{\pm} 
= 2\hbar\omega_0 \zeta +\frac{E_{RI}+E_{W_{RI}}\pm (E_{CE}+E_{W_{CE}})}{1\pm\Omega^2},
\end{equation}
where 
\begin{eqnarray*}
E_{RI} &=& \hbar\omega_0 c_s \sqrt{\zeta}\exp{\left[-\zeta d^2\right]}I_0(\zeta d^2),\\
E_{W_{RI}} &=& \hbar\omega_0 \left[2d^2(1-\mbox{Erf}(d\sqrt{\zeta }))
-\frac{2d}{\sqrt{\zeta \pi}}\exp{[-\zeta d^2]}\right],\\
E_{CE} &=&  \hbar\omega_0 c_s \sqrt{\zeta }
\exp{\left[-\zeta d^2(2+\vartheta^2)\right]}I_0(\zeta d^2\vartheta^2),\\
E_{W_{CE}} &=& -\hbar\omega_0\frac{2d}{\sqrt{\pi \zeta }}\exp{[-2\zeta d^2(1+\vartheta^2)]}.
\end{eqnarray*}
The components of the vectors $\mathbf{a^\prime}$ and $\mathbf{b^\prime}$ are
\begin{eqnarray}
a_x' &=& 0,\;\;\;\;\;\;
a_y' = 0,\\
b_x' &=& -\frac{\hbar^2}{2m l_{d}}\frac{\Omega^2}{\sqrt{1-\Omega^4}} \frac{\zeta  d}{l_0}(1-\vartheta^2),\\
b_y' &=& -\frac{\hbar^2}{2m l_{br}}\frac{\Omega^2}{\sqrt{1-\Omega^4}} \frac{\zeta  d}{l_0}(1-\vartheta^2).
\end{eqnarray}
where $l_{br}$ and $l_{d}$ are the spin-orbit lengths for the Rashba
and Dresselhaus respectively.

The matrix elements of the vector $\bf{n}$ are
\begin{eqnarray}
\langle\Psi_+|n_{x,1}|\Psi_-\rangle &=& -\frac{dl_0}{\sqrt{1-\Omega^4}}
\left(\frac{1}{l_d}+{\rm i}\Omega^2\frac{\vartheta}{l_{br}}\right),\\
\langle\Psi_+|n_{y,1}|\Psi_-\rangle &=& -\frac{dl_0}{\sqrt{1-\Omega^4}}
\left(\frac{1}{l_{br}}+{\rm i}\Omega^2\frac{\vartheta}{l_d}\right),
\end{eqnarray}
and
\begin{equation}
 \mu B_{so} = \frac{K_-}{1-\Omega^2} \vartheta \left[1-\Omega^2
(1-\zeta d-\zeta d^2\vartheta^2)\right].
\end{equation}

\section{Spin matrices}\label{spinmatrices}
In the singlet and triplet basis, one can evaluate the sixteen  matrices 
which can be formed as
the direct product of two Pauli matrices and the identity. Here we list 
only the matrices
needed for our purposes, and we regroup them to
combinations in which they appear in the text.

\begin{equation}
\boldsymbol{\sigma}_{1}\cdot\boldsymbol{\sigma}_{2}=\left(
\begin{array}{cccc}
-3 & 0 & 0 & 0\\
0 & 1 & 0 & 0\\
0 & 0 & 1 & 0\\
0 & 0 & 0 & 1
\end{array}
\right), 
\end{equation}

\begin{widetext}
\begin{equation}
\boldsymbol{\sigma}_{1} - \boldsymbol{\sigma}_{2}=
\Big\{
\left(
\begin{array}{cccc}
0 & 0 & -\sqrt{2} & \sqrt{2}\\
0 & 0 & 0 & 0\\
-\sqrt{2} & 0 & 0 & 0\\
\sqrt{2} & 0 & 0 & 0
\end{array}
\right),
\left(
\begin{array}{cccc}
0 & 0 & -\sqrt{2}{\rm i} & -\sqrt{2}{\rm i}\\
0 & 0 & 0 & 0\\
\sqrt{2}{\rm i} & 0 & 0 & 0\\
\sqrt{2}{\rm i} & 0 & 0 & 0
\end{array}
\right),
\left(
\begin{array}{cccc}
0 & 2 & 0 & 0\\
2 & 0 & 0 & 0\\
0 & 0 & 0 & 0\\
0 & 0 & 0 & 0
\end{array}
\right)\Big\} ,
\end{equation}

\begin{equation}
\boldsymbol{\sigma}_{1}\times\boldsymbol{\sigma}_{2}=
\Big\{
\left(
\begin{array}{cccc}
0 & 0 & -\sqrt{2}{\rm i} & \sqrt{2}{\rm i}\\
0 & 0 & 0 & 0\\
\sqrt{2}{\rm i} & 0 & 0 & 0\\
-\sqrt{2}{\rm i} & 0 & 0 & 0
\end{array}
\right),
\left(
\begin{array}{cccc}
0 & 0 & \sqrt{2} & \sqrt{2}\\
0 & 0 & 0 & 0\\
\sqrt{2} & 0 & 0 & 0\\
\sqrt{2} & 0 & 0 & 0
\end{array}
\right),
\left(
\begin{array}{cccc}
0 & 2{\rm i} & 0 & 0\\
-2{\rm i} & 0 & 0 & 0\\
0 & 0 & 0 & 0\\
0 & 0 & 0 & 0
\end{array}
\right)\Big\},
\end{equation}

\begin{equation}
\boldsymbol{\sigma}_{1} + \boldsymbol{\sigma}_{2}=
\Big\{
\left(
\begin{array}{cccc}
0 & 0 & 0 & 0\\
0 & 0 & \sqrt{2} & \sqrt{2}\\
0 & \sqrt{2} & 0 & 0\\
0 & \sqrt{2} & 0 & 0
\end{array}
\right),
\left(
\begin{array}{cccc}
0 & 0 & 0 & 0\\
0 & 0 & \sqrt{2}{\rm i} & -\sqrt{2}{\rm i}\\
0 & -\sqrt{2}{\rm i} & 0 & 0\\
0 & \sqrt{2}{\rm i} & 0 & 0
\end{array}
\right),
\left(
\begin{array}{cccc}
0 & 0 & 0 & 0\\
0 & 0 & 0 & 0\\
0 & 0 & 2 & 0\\
0 & 0 & 0 & -2
\end{array}
\right)\Big\}.
\end{equation}
\end{widetext}


\end{document}